\documentstyle[12pt]{article}

\newcommand{\be}{\begin{equation}}
\newcommand{\ee}{\end{equation}}
\newcommand{\bea}{\begin{eqnarray}}
\newcommand{\eea}{\end{eqnarray}}
\newcommand{\beaa}{\begin{eqnarray*}}
\newcommand{\eeaa}{\end{eqnarray*}}

\newcommand{\hsp}{\hspace{0.1in}}

\newcommand{\del}{\partial}
\newcommand{\g}{{\cal G}}
\newcommand{\J}{{\cal J}}
\newcommand{\K}{{\cal K}}
\newcommand{\p}{{\cal P}}
\newcommand{\Lg}{{\bf g}}
\newcommand{\Qt}{ {\tilde Q} }
\advance\textheight by \topskip
\makeatletter

\@addtoreset{equation}{section}
\def\section{\@startsection {section}{1}{\z@}{-3.5ex plus -1ex minus
 -.2ex}{2.3ex plus .2ex}{\large\bf\centering}}
\def\subsection{\@startsection{subsection}{2}{\z@}{-3.25ex plus%
 -1ex minus -.2ex}{1.5ex plus .2ex}{\bf}}
\def\subsubsection{\@startsection{subsubsection}{3}{\z@}{-3.25ex plus%
 -1ex minus -.2ex}{1.5ex plus .2ex}{\sl}}
\makeatother

\begin{document}

\baselineskip 18pt
\parindent 12pt
\parskip 10pt
{
\parskip 0pt
\newpage
\begin{titlepage}
\begin{flushright}
DAMTP-98-171\\
Imperial/TP/98-99/21\\
hep-th/9902008v2\\
January 1999\\[3mm]
\end{flushright}
\vspace{.4cm}
\begin{center}
{\Large {\bf
Local conserved charges in principal chiral models}}\\
\vspace{1cm}
{\large J.M. Evans${}^a$\footnote{e-mail: J.M.Evans@damtp.cam.ac.uk}, 
M. Hassan${}^a$\footnote{e-mail: M.U.Hassan@damtp.cam.ac.uk},
N.J. MacKay${}^b$\footnote{e-mail: N.MacKay@sheffield.ac.uk}, 
A.J. Mountain${}^c$\footnote{e-mail: A.Mountain@ic.ac.uk}}
\\
\vspace{3mm}
{\em ${}^a$ DAMTP, University of Cambridge, Silver Street, Cambridge
CB3 9EW, U.K.}\\
{\em ${}^b$ Department of Applied Mathematics, 
University of Sheffield, Sheffield S3 7RH, U.K.}\\
{\em ${}^c$ Blackett Laboratory, Imperial College, Prince Consort Road,
London SW7 2BZ, U.K.}\\

\vspace{1.5cm}
{\bf{ABSTRACT}}
\end{center}
\begin{quote}{\small
Local conserved charges in principal chiral models 
in 1+1 dimensions are investigated.
There is a classically conserved local charge for each totally
symmetric invariant tensor of the underlying group. 
These local charges are shown to be in involution with 
the non-local Yangian charges.
The Poisson bracket algebra of the local charges is then studied.
For each classical algebra, an infinite set of
local charges with spins equal to the exponents modulo the 
Coxeter number is constructed, and it is shown 
that these commute with one another.
Brief comments are made on the evidence for, and implications of, 
survival of these charges in the quantum theory.
}
\end{quote}
\vfill
\end{titlepage}
}

\section{Introduction}

Integrable lagrangian field theories in 1+1 dimensions exhibit
various kinds of higher-spin conserved quantities which 
constrain their quantum behaviour, forcing their S-matrices to
factorize and allowing them to be determined exactly in many instances.
Such exotic symmetries can usually be traced to underlying 
mathematical structures which incorporate Lie algebras in some way.
Beyond these broad similarities, however, one encounters many
examples with profound physical and mathematical differences. 
One of the most important distinctions is between
conserved charges which are integrals of {\it local\/} 
functions of the fields, and those which instead involve 
{\it non-local\/} functions of the fields. 

Local charges have been studied extensively in affine Toda
theories (ATFTs) for which the defining Lie algebra data are just a 
set of simple roots (plus the lowest root).
Each ATFT possesses infinitely many commuting local charges with 
spins equal to the exponents of
the Lie algebra modulo its Coxeter number $h$ \cite{wils80}:
\begin{eqnarray}
\label{exps}
a_\ell= su(\ell\!+\!1) & 1,2,3,\ldots,\ell & h=\ell\!+\!1\nonumber\\
b_\ell= so(2\ell\!+\!1) & 1,3,5,\ldots,2\ell\!-\!1 & h=2\ell \nonumber\\
c_\ell= sp(2\ell) & 1,3,5,\ldots,2\ell\!-\!1  & h=2\ell\nonumber\\
d_\ell= so(2\ell) & \hspace{0.2in}1,3,5,\ldots,2\ell\!-\!3\,,\,\ell\!-\!1 
\hspace{0.2in} &h=2\ell\!-\!2
\end{eqnarray}
Their local nature means that these charges are 
additive on asymptotic, multi-particle states, and the fact that they
commute means that single-particle states 
may be chosen to be simultaneous eigenstates of them.
Their existence places strong constraints on the possible 
three-point couplings, and their relationship with the underlying
Lie algebra in the theory is crucial in ensuring that there is a
consistent solution of the bootstrap equations 
(see \cite{brade90} and {\it e.g.}~\cite{corri94}) as 
was made clear in the elegant construction
of Dorey \cite{dorey91}.

Non-local charges have proved very important in
understanding the classical and quantum integrability of 
certain non-linear sigma-models \cite{lusch78,brez79}, 
amongst which are the 
principal chiral models (PCMs) with target space some compact 
Lie group. They also occur in ATFTs with {\it imaginary\/} coupling,
in connection with the soliton solutions in these theories.
In marked distinction to local charges, non-local charges 
are not additive on multi-particle states and they can have either 
indefinite or non-integral spin.
The non-local charges generate a quantum group structure 
which underpins the factorizability of the S-matrix by 
naturally providing solutions of the Yang-Baxter equation
(see {\it e.g.}~\cite{ganden96}).
The quantum group is a Yangian \cite{berna91} 
in the case of PCMs, whereas it is a quantized affine algebra 
for imaginary coupling ATFTs \cite{berna90}. 

There are some very important physical differences between the types
of models we have just mentioned. 
In ATFTs the coupling constant is small at low energies, 
so that perturbation theory and semi-classical techniques can be used 
to determine the spectrum of particles (including solitons for the
imaginary coupling theories). 
One-loop calculations confirm that 
mass ratios of particles are unchanged for the simply-laced algebras, 
while they vary with the coupling constant for the nonsimply-laced cases 
(see {\it e.g.}~\cite{corri94,brade90}).
In complete contrast, any sigma-model with a compact 
target manifold---in particular a PCM---is strongly-coupled in 
the infra-red, so that information about
the quantum theory is much harder to extract from the lagrangian.
In fact the classical lagrangians for these theories are 
scale-invariant, and masses arise through a 
complicated quantum-mechanical effect at strong coupling.

Despite the disparate properties 
of these field theories and the very different roles played by
the local and non-local charges within them, it turns out that they 
still have much in common at the quantum level, and in particular
the patterns of masses and three-point couplings
coincide in a rather remarkable way.
The mass ratios which emerge for PCMs (from their exact S-matrices
\cite{ogie86,poly83,abda84}) 
are actually identical to those of the 
ATFTs \cite{corri94,brade90} 
if one considers theories based on the same, simply-laced Lie 
algebra. For nonsimply-laced algebras there is a more subtle connection, 
in which the PCM mass ratios coincide with those of
tree-level ATFTs based on twisted affine algebras.

The three-point couplings in ATFTs are given by Dorey's rule,
while the three-point couplings in PCMs arise as 
S-matrix fusions contained in the tensor product rule for 
representations of the Yangian (the analogue of the 
Clebsch-Gordan rule for Lie algebras). 
There is no reason to suppose {\it a priori\/} that these should be
related, and yet it has recently been proved \cite{chari95} that 
the quantum group fusion rule for fundamental representations 
(derived from the non-local charge algebra) 
and Dorey's rule (derived from local charges) are one and the same.
Since the proof consists of a complicated case-by-case analysis,
however, it remains something of a mystery why this is so.

Our aim in this paper is to study in detail the properties of 
certain local conserved charges in the principal chiral models.
Their existence has been known for some time \cite{gold80,aref78},
but they have been somewhat neglected in PCMs in favour 
of the non-local charges discussed above.
An obvious possibility is that the PCM local charges can  
offer a more transparent explanation for the striking
common features of PCMs and ATFTs.
Another motivation is that the detailed study of this particular
model in which the local and non-local charges co-exist
may offer some insights into the way in which these entities
are related in general.

The majority of the paper---sections 2, 3 and 4---elucidates 
the classical properties of the local charges appearing in PCMs.
There exists a conserved local charge for each
invariant tensor or Casimir of the underlying Lie algebra, and we
shall prove that these always commute with the non-local Yangian charges.
We then investigate in detail the Poisson bracket algebra
of the local charges with one another.
Our main result is to find for each PCM an infinite family of 
local charges which commute classically and which have spins
equal to the exponents (\ref{exps}) modulo the Coxeter number, 
precisely as for the ATFTs. 
Because of the difficulties inherent in quantizing the PCM, 
there is only a limited amount which can be said about the
quantum behaviour of these charges, which we summarize in section 5.
We can, nevertheless, examine the implications of their survival 
at the quantum level, and we find 
complete compatibility with the known multiplet structures and 
exact S-matrices \cite{ogie86,abda84}.

\section{The classical principal chiral model}

\subsection{The lagrangian, symmetries and currents}

The principal chiral model (PCM) is defined by the lagrangian 
\be\label{pcmlagr}
{\cal L} = {\kappa \over 2 } \, {\rm Tr}\left( \partial_\mu g^{-1}
 \partial^\mu g\right) 
\ee
where the field $g(x^\mu)$ takes values in some compact Lie group ${\cal G}$
and $\kappa$ is a dimensionless coupling which may be set to any
convenient value in the classical theory, without loss of generality.
There is a global continuous symmetry 
\be\label{chsym}
{\cal G}_L \times {\cal G}_R \; : \qquad 
g \mapsto U^{\phantom{1}}_{\rm L} \,g\, U^{-1}_{\rm R}
\ee
with corresponding conserved currents 
\be\label{lrcurr} 
j_\mu^L=\kappa \, \partial_{\mu} g \,g^{-1} , \qquad 
j_\mu^R = - \kappa \, g^{-1}\partial_{\mu} g
\ee
which take values in the Lie algebra $\Lg$ of ${\cal G}$.
The equations of motion following from (\ref{pcmlagr}) 
correspond to the conservation of these currents:
\be\label{cons}
\partial^{\mu}j_{\mu}(x,t)=0 \, . 
\ee
They also obey 
\be\label{cf}
\partial_{\mu} j_{\nu} - \partial_{\nu} j_{\mu} - {1 \over \kappa} 
[j_{\mu},j_{\nu}] = 0 
\ee
identically, as a consequence of their definitions in terms of the field $g$.
Here and elsewhere we will adopt the convention that any equation
written for a current $j_\mu$ without a label holds true for both 
$L$ and $R$ currents.
It is significant that the last condition above can be interpreted as a
zero-curvature condition for a connection with covariant
derivative
\be 
\label{covderiv}
\nabla_\mu X = \del_\mu X - {1 \over \kappa} [j_\mu , X]
\ee
acting on any $X$ in $\Lg$.
Here we view $- \kappa^{-1} j_\mu$ as a two-dimensional
gauge field, its definition in terms of $g$ implying that 
it is pure-gauge.
The two conditions (\ref{cons},\,\ref{cf}) capture 
the entire algebraic structure of the PCM.

We shall restrict attention to 
the classical groups $\g = SU(\ell)$, $SO(\ell)$, $Sp(\ell)$ (with
$\ell$ even in the last case)
with the field $g(x^\mu)$ in the defining representation.
The corresponding Lie algebra $\Lg$ then consists
of $\ell{\times}\ell$ complex matrices $X$ which obey
\bea 
su(\ell): &\quad & X^\dag = - X, \quad {\rm Tr}(X) = 0\nonumber \\
so(\ell): &\quad & X^* = X, \quad \quad X^T = - X \label{SOUP} \\
sp(\ell): &\quad & X^\dag = -X, \quad X^T = - JXJ^{-1} \nonumber
\eea
where $J$ is some chosen symplectic structure.
In each case we introduce a basis of anti-hermitian generators $t^a$ for 
$\Lg$ with real structure constants $f^{abc}$ and
normalizations given by 
\be 
[ t^a , t^b ] = f^{abc} t^c \; , \qquad {\rm Tr} (t^a t^b) = -
\delta^{ab} \; .
\ee 
(Lie algebra indices will always be taken from the beginning of the
alphabet.)
For any $X \in \Lg$ we write 
\be\label{alcpts}
X = t^{a} X^a \;  \qquad X^a = - {\rm Tr}(t^a X) \; .
\ee

Spacetime symmetries will also play an important role in what follows.
The classical PCM lagrangian is conformally-invariant, and as a result the
energy momentum tensor
\be 
T_{\mu \nu} = -{1 \over 2 \kappa} \left ( \, 
{\rm Tr} (j_\mu j_\nu) - {1\over2}\eta_{\mu \nu} {\rm Tr}
 (j_\rho j^\rho) \, \right )
\ee
is not only conserved and symmetric but also traceless.
We shall use standard orthonormal coordinates
$x^0 = t$ and $x^1 = x$ in two dimensions, 
as well as light-cone coordinates and their derivatives defined by
$$
x^\pm = {1\over 2} (t \pm x) \, , \qquad \del_\pm = \del_t \pm \del_x
\, .
$$ 
The equations (\ref{cons},\,\ref{cf}) can then be written 
\be\label{lce}
\partial_- j_+ = - \partial_+ j_- = - {1\over 2\kappa} [j_+,j_-] \; ,
\ee
whilst the energy-momentum tensor takes the familiar form 
\be 
T_{\pm \pm} = -{1 \over 2 \kappa} 
{\rm Tr} ( j_{\pm} j_{\pm}) \; , \qquad T_{+-} = T_{-+} = 0
\; ,
\ee
with 
\be\label{conf}
\partial_- T_{++} = \partial_+ T_{--} = 0 \; .
\ee

In addition to the continuous symmetries comprising $\g_L$ and $\g_R$,
there are also important discrete symmetries of the principal 
chiral model. 
For any PCM there is a symmetry 
\be\label{parity}
\pi \, : \; g\mapsto g^{-1} \qquad \Rightarrow
 \qquad j^L \leftrightarrow j^R \, 
\ee
which exchanges $\g_L$ and $\g_R$ and which we shall therefore refer
to as $\g$-parity. (In the PCM effective field theory
description of strong interactions in four dimensions, 
the physical parity operator 
is our $\g$-parity together with spatial reflection.) 
Additional discrete symmetries arise as outer automorphisms of $\g$ 
acting on the the field $g$.
Thus we have 
\be 
\qquad \gamma\,:\;g\mapsto g^*  \qquad \Rightarrow \qquad
j^L \mapsto (j^L)^* = -(j^L)^T \;, \qquad
j^R \mapsto (j^R)^* = -(j^R)^T \,,
\ee
which exchanges complex-conjugate representations.
This realizes the outer automorphism of $\Lg = su(\ell)$.
The map is trivial (up to conjugation) 
if $\Lg$ has only real (or pseudo-real) representations. 
For $\Lg = so(2 \ell)$ we also have
$$
\qquad\sigma\,:\;g \mapsto MgM^{-1}\qquad  \Rightarrow \qquad
j^L \mapsto M j^L M^{-1} \;, \qquad
j^R \mapsto M j^R M^{-1} \,,
$$
where $M$ is a fixed matrix with determinant $-1$. 
This is the outer automorphism which exchanges the inequivalent 
spinor representations.
The maps $\sigma$ and $\gamma$ coincide (up to conjugation) 
for $\Lg = so(2\ell)$ when $\ell$ is odd.

\subsection{Canonical formalism}

The canonical Poisson brackets for the theory are 
\begin{eqnarray}
\left\{ j_0^a (x), j_0^b (y) \right\} 
& = & f^{abc} \, j_0^c (x) \, \delta(x{-}y) \nonumber \\
\left\{ j_0^a (x), j_1^b (y) \right\} 
& = & f^{abc} \, j_1^c (x) \, \delta(x{-}y) + \kappa \, \delta^{ab} 
\, \delta'(x{-}y) \label{PBs}\\
\left\{ j_1^a (x), j_1^b (y) \right\} 
& = & 0 \nonumber
\end{eqnarray}
at equal time.
These expressions hold for {\em either\/} of the currents $j^L$ or $j^R$
separately, while the algebra of $j^L$ with $j^R$
involves only $\delta'(x{-}y)$ terms in the brackets 
of space- with time-components 
\cite{fadd87} (we shall not need them here) 
in keeping with the direct product structure of $\g_L \times \g_R$.
For light-cone current components these brackets become 
\begin{eqnarray}
\left\{ j_\pm^a (x), j_\pm^b (y) \right\} 
& = & f^{abc} ( \, {\textstyle{3 \over 2}} j_\pm^c (x) - 
{\textstyle {1 \over 2}} j_\mp^c (x) \, ) \,
\delta(x{-}y) \, \pm \, 2 \kappa \, \delta^{ab} \, \delta'(x{-}y) \nonumber \\
\left\{ j_+^a (x), j_-^b (y) \right\} 
& = & {\textstyle {1\over 2}}f^{abc} (\, j_+^c(x)+j_-^c(x) \,) \, 
\delta(x{-}y) 
\label{LCPBs} 
\end{eqnarray}
They imply that the energy momentum tensor satisfies the 
classical, centre-less Virasoro algebra 
\be\label{vir}
\{ T_{\pm\pm} (x) , T_{\pm\pm} (y) \} = \pm 2 
( \, T_{\pm\pm}(x) + T_{\pm\pm} (y) \, )  \, \delta'(x{-}y) \; .
\ee

There are a number of ways to derive the expressions (\ref{PBs}). 
One approach \cite{fadd87} has the advantage of dealing directly 
with the currents rather than with the underlying field $g$: 
we can select $j_1$ as the only independent dynamical variable, 
since we can regard $j_0$ as a function of it which is determined through
the relation (\ref{cf}).
The price to be paid is the introduction of an operator
$\nabla_1^{-1}$ which is
non-local in space and whose definition assumes suitable boundary 
conditions for the currents at spatial infinity.
This enables us to write $j_0 = \nabla_1^{-1} ({\del_0 j_1})$
and so write the lagrangian as (we set $\kappa =1$ here for simplicity)
$$
{\cal L} = - {1\over 2} {\rm Tr} (j_0^2 - j_1^2) 
= - {1\over 2} {\rm Tr} 
\left [ \, ( \nabla^{-1}_1 {\del_0 j_1 } )^2  - j_1^2 \right
]
$$
The momentum conjugate to $j_1$ is defined in the usual way:
$\pi_1 = \del {\cal L} / {\del (\del_0 j_1)} = - \nabla^{-2}_1 \del_0
j_1$ and we deduce that $j_0 = -\nabla_1 \pi_1$.
(This requires the property $\nabla_1^{-1}(A) B = - A \nabla^{-1}_1(B)$ up
to terms which vanish on integrating over space, so again 
the adoption of suitable boundary conditions on the fields is crucial.) 
The expressions (\ref{PBs}) can now be recovered 
from the standard equal-time Poisson brackets for $j_1$ and $\pi_1$
after a short calculation.
A lengthier but more routine derivation which avoids the use of the operator
$\nabla_1^{-1}$ can be found in an appendix.

\section{Classical conserved charges}

\subsection{Non-local charges}
There exist infinitely many conserved, Lie algebra-valued, 
non-local charges in the PCM, which generate a Yangian 
$Y(\Lg)$ \cite{mack92}. 
In fact there are two copies of this structure, 
constructed either from $j_\mu^L$ or $j_\mu^R$, 
and so the model has a charge algebra $Y_L(\Lg)\times Y_R(\Lg)$. 
(It can be checked that $Y_L$ and $Y_R$ commute.) 
A full set of non-local charges $Q^{(n)a}$ with $n=0,1,2,\ldots,$ 
can be generated from  the obvious local charge 
\be\label{qzero} 
Q^{(0)a}  =  \int^{\infty}_{-\infty} j_{0}^{a} {dx} 
\ee 
and the first non-local charge
\be\label{qone} 
Q^{(1)a}  =  \int^{\infty}_{-\infty} j_{1}^{a} {dx} - 
{1\over 2\kappa}f^{abc}\int^{\infty}_{-\infty} j_{0}^{b}(x) 
\int_{-\infty}^{x} j_{0}^{c}(y) \,{dy} \,{dx} \,.
\ee 
This set can also be defined 
by a power series expansion of the transfer matrix in its 
spectral parameter, or equivalently
they can be constructed by the following iterative procedure \cite{brez79}.

Suppose we have Lie algebra-valued 
currents $j_{\mu}^{(r)}$ defined for $r=0, 1, \ldots , n$ 
which are conserved:
\be
\del_- j^{(r)}_+ + \del_+ j^{(r)}_- = 0 \quad \Longleftrightarrow \quad
j_\pm^{(r)}=\pm\partial_\pm{\chi}^{(r)}
\ee 
for some scalar Lie algebra-valued 
functions ${\chi}^{(r)}$, and that these currents are related to one another 
by 
\be 
j_{\mu}^{(r+1)} = \nabla_{\mu} {\chi}^{(r)} =
\partial_{\mu} \chi^{(r)} - \kappa^{-1} [j_{\mu} , {\chi}^{(r)} ] \, .
\ee
Taking $r=n$ defines a new current, $j_\mu^{(n+1)}$
which is conserved because
\beaa
\del _- j_+^{(n+1)} + \del_+ j_-^{(n+1)} 
&=& (\del_- \nabla_+ + \del_+ \nabla_- ) \chi^{(n)} \\
&=& (\nabla_+ \del_- + \nabla_- \del_+ ) \chi^{(n)} \\
&=& - \nabla_+ j_-^{(n)} + \nabla_- j_+^{(n)} \\
&=& - [ \nabla_+ , \nabla_- ] \chi^{(n-1)} \\
&=& 0 
\eeaa
for $n \geq 1$. Thus, starting from $j_\mu^{(0)} = j_\mu$ and
$j_\mu^{(1)}$ given by
$$
j_0^{(1)a}(x) = j_{1}^{a}(x) - {1\over 2\kappa}f^{abc} j_{0}^{b}(x) 
\int_{-\infty}^{x} j_{0}^{c}(y) \,dy \; ,
\qquad j_1^{(1)a}(x) = j_{0}^{a}(x) - {1\over 2\kappa}f^{abc} j_{1}^{b}(x) 
\int_{-\infty}^{x} j_{1}^{c}(y) \, dy
$$
we can define an infinite set of currents which are 
non-local functions of the original fields for $n>0$;
their conserved charges are $Q^{(n)a} = \int j_0^{(n)a} dx $.

Classically, the non-local charges are Lorentz scalars: 
applying the boost operator $M$ we obtain 
$ 
\left\{ M, Q^{(0)a} \right\}  =
 \left\{ M, Q^{(1)a} \right\} =0 .
$
(The second of these equations is modified in the quantum theory,
however---see below.)
Because the charges are non-local they will not generally be additive
on products of states. Their action is 
given by the coproduct: 
\begin{eqnarray}
\Delta \left( Q^{(0)a} \right) & = & Q^{(0)a} \otimes 1 + 
1 \otimes Q^{(0)a} \nonumber \\
{\rm and} \hspace{0.2in}\Delta \left( Q^{(1)a} \right) & = 
& Q^{(1)a} \otimes 1 + 1 \otimes Q^{(1)a} + {1\over2\kappa} 
f^{abc} Q^{(0)b} \otimes Q^{(0)c} \hsp,
\end{eqnarray}
which we see is non-trivial in the second case.
As well as the usual interpretation in the quantum theory,
these equations may also be interpreted classically as giving the 
values of the charges on widely-separated, localized field 
configurations \cite{lusch78}.

\subsection{Local charges}
In any conformally-invariant theory, 
the conservation of the energy-momentum tensor (\ref{conf}) 
immediately implies a series of higher-spin conservation 
laws:
\be\label{confpower}
\partial_- (T^n_{++}) = \partial_+ (T_{--}^n) = 0 \, . 
\ee
But the PCM has more basic conservation laws
which depend on the detailed form of the equations of motion of the
currents rather than on conformal invariance alone.
The simplest examples are 
\be\label{curr}
\partial_- {\rm Tr} (j_{+}^m) = 
\partial_+ {\rm Tr} (j_{-}^m) = 0
\ee
which follow easily from (\ref{lce}).
More generally, we may consider any rank-$m$, totally
symmetric, invariant tensor $d_{a_1 a_2 \ldots a_m}$ associated 
with a Casimir operator 
\be\label{cas}
{\cal C}_m = d_{a_1 a_2 \ldots a_m} t^{a_1} t^{a_2} \ldots t^{a_m}
\ee
where 
\be\label{dinv} 
[ {\cal C}_m , t_b ] = 0 \qquad \iff \qquad
d_{c(a_1a_2 \ldots a_{m-1}}f_{a_m)bc} = 0 \,
\ee
(and as usual $( \ldots)$ denotes symmetrization of the enclosed indices).
It is then easy to check that invariance of $d$ 
ensures the conservation equations
\be 
\label{gencons}
\del_\mp( \, d_{a_1 a_2 \ldots a_m} j_\pm^{a_1} j_\pm^{a_2} \ldots
j_\pm^{a_m} \, ) = 0 \ .
\ee
The corresponding conserved charges will be denoted
\be \label{locch}
q_{\pm s} \,=\, \int_{-\infty}^\infty d_{a_1a_2 \ldots a_m} \,
j_\pm^{a_1}(x) j_\pm^{a_2} (x) \ldots j_\pm^{a_m}(x)
\,dx 
\ee
and labelled by their spin $s= m{-}1$ (the Poisson bracket with the
boost generator $M$ is $\{ M , q_{\pm s} \} = \pm s q_{\pm s}$).
We shall refer to charges $q_{\pm s}$ with $s>0$ 
as having positive/negative chirality.

Invariance of the $d$-tensor implies 
that the same local conservation laws are obtained using 
either of the currents $j^L$ or $j^R$, so there is just a single copy 
of these local charges, unlike the 
two-fold $L$ and $R$ copies of the non-local charges.
Also in contrast to the non-local charges, we note that any local charge
must be additive on multi-particle states, which we can also express
by saying that such a charge has a trivial co-product:
$\Delta(q_s) = q_s \otimes 1 + 1 \otimes q_s $.

The currents in (\ref{confpower}) correspond to  
even-rank invariant tensors constructed from Kronecker deltas:
\be \label{sdelta}
d_{a_1a_2 \ldots a_{2n-1}a_{2n}} = 
\delta_{(a_1 a_2} \delta_{a_3 a_4} \! \ldots \delta_{a_{2n-1} a_{2n} )} 
\ee
while those in (\ref{curr}) correspond to 
\be \label{strace}
d_{a_1a_2 \ldots a_m} = 
{\rm STr}(t^{a_1} t^{a_2} \! \ldots t^{a_m}) 
\ee
with `STr' denoting the trace of a completely symmetrized product of matrices.
For $su(\ell)$ this tensor is non-vanishing for each integer
$m$, but for $so(\ell)$ or $sp(\ell)$ it is non-zero only when
$m$ is even. 
It is useful to introduce the notation 
\be\label{simple}
\J_m = {\rm Tr} (j_+^m) 
\ee
for the corresponding currents.
Notice that $\J_2$ is proportional to the energy-momentum tensor
$T_{++}$, so that the currents in (\ref{confpower}) can also be 
written $(\J_2)^n$.
 
There are infinitely many invariant tensors $d_{a_1 \ldots a_m}$ 
for each algebra $\Lg$, but there are only
$rank(\Lg)$ independent or {\it primitive\/} $d$-tensors and
Casimirs (see {\it e.g.}~\cite{azca97}) with degrees equal to the
exponents of $\Lg$ plus one.
All other invariant tensors can be expressed as polynomials in these and 
the structure constants $f_{abc}$. 
The choice of these $rank (\Lg)$ primitive tensors is not unique, 
the ambiguity being the addition of polynomials in tensors
of lower rank.
The symmetrized traces in (\ref{strace}) are a particular 
choice for all the primitive $d$-tensors of the 
classical algebras, with one exception.
This exception is the Pfaffian invariant in $so(2\ell)$, which has rank
$\ell$ and can be written 
\be\label{pfaff}
d_{a_1...a_\ell} = \epsilon_{i_1j_1 \ldots i_\ell j_\ell} 
(t^{a_1})_{i_1 j_1} \ldots
(t^{a_\ell})_{i_\ell j_\ell} \; .
\ee
This tensor cannot be expressed as a trace in the defining representation,
although it is related to a trace in the spinor representation.
We denote the corresponding current by
\be
\label{pfaffcurr}
\p_{\ell} = \epsilon_{i_1 j_1 \ldots i_\ell j_\ell} (j_{+})_{i_1 j_1} \ldots
(j_{+})_{i_\ell j_\ell} \; .
\ee

Finally, we should mention that there are infinitely many 
more conserved currents in the PCM which arise as 
{\it differential\/} polynomials in those already discussed. 
For example, $\del_- \left ( {\rm Tr} (j_+^p ) \del_+^r {\rm Tr}
(j_+^q) \right ) = 0$ follows immediately from (\ref{curr}).
We shall not be directly concerned with the properties of these 
more general currents.  

\subsection{Commutation of local with non-local charges}
We will now show that all local charges of 
the general type (\ref{locch})  
commute with the non-local charges generated by
$Q^{(0)a}$ and $Q^{(1)a}$.  This means showing that 
\be\label{lnlcr}
\{\, q_s, \, Q^{(0)b} \, \} = \{ \, q_s, \, Q^{(1)b}\,\}= 0 \; .
\ee
The vanishing of the first bracket follows immediately from 
invariance of the $d$-tensor used to define $q_s$; this 
says simply that the charge $q_s$ is a singlet under the Lie algebra.
The calculation of the second bracket is more delicate,
and involves a cancellation between contributions originating 
from ultralocal and non-ultralocal terms. 

Consider the expression (\ref{qone}) for $Q^{(1)b}$, which involves
two terms. The bracket of $q_s$ with the first (local) term is
\begin{eqnarray} 
\{\, q_s, \int\!dy \,j_1^b(y) \, \} & = &  
d_{a_1a_2...a_m} \int \int dx\,dy\, \{\, j_+^{a_1}(x)\ldots j_+^{a_m}(x),
\, j_1^b(y) \, \} \nonumber \\[4pt]
& = & -m \, d_{a_1a_2...a_{m-1}c} f^{bcd}
\int\!dx\, 
j_+^{a_1}(x) \ldots j_+^{a_{m-1}}(x) j_1^d(x) \, .
\label{term1}
\end{eqnarray}
We have used the fact that $d_{a_1 \ldots a_m}$ is totally symmetric
and also the boundary conditions $j\rightarrow 0$ as
$x\rightarrow\pm\infty$ which imply that there is no contribution from
$\delta'$ terms in the current Poisson bracket.
To deal with the second (non-local) term in (\ref{qone})
we must handle carefully the limits 
of the spatial integration: we take these to be $\pm L$ and only
afterwards let $L\rightarrow \infty$.
We must calculate 
\beaa
&&\mbox{\hskip -15pt}  
\{ \, q_s \, ,  
\int_{-L}^L \! dy \int_{-L}^y \!dz\,f^{bcd} j_0^c(y)
j_0^d(z) \, \, \} 
\\[2pt]
&&\mbox{\hskip 5pt}
= \; 
m \, d_{a_1a_2...a_m} f^{bcd}
\int_{-L}^L \! dx \int_{-L}^L \! dy \int_{-L}^y \!dz\,
j_+^{a_1}(x)\ldots j_+^{a_{m-1}}(x)\, \{ j_+^{a_m} (x) , \, j_0^c(y)
j_0^d(z)  \, \} 
\\[2pt]
&&\mbox{\hskip -15pt}
= \; 
\kappa m \, d_{a_1a_2...a_{m-1}c} f^{bcd}
\int_{-L}^L \! dx \int_{-L}^L \! dy \int_{-L}^y \!dz\,
j_+^{a_1}(x)\ldots j_+^{a_{m-1}}(x)
\, [ \, \, j_0^d(z) \, \delta'(x{-}y) 
\, - \, j_0^d(y) \, \delta'(x{-}z) \, \, ]  
\eeaa
from (\ref{PBs}). Introducing a step-function, the 
multiple integral can be written
\beaa
&&\mbox{\hskip -10pt}  
\int_{-L}^L \int_{-L}^L \int_{-L}^L \! dx \, dy \, dz \,
j_+^{a_1}(x)\ldots j_+^{a_{m-1}}(x) \, j_0^d(z) \, 
[ \, \theta(y{-}z) - \theta(z{-}y) \, ] \, \delta'(x{-}y) 
\\[4pt]
&&
= 
\int_{-L}^L \int_{-L}^L \! dx \, dz\,
j_+^{a_1}(x)\ldots j_+^{a_{m-1}}(x) \, j_0^d(z) \,
[ \, \delta(x{-}z) - \delta(x{+}L) -\delta(x{-}L) + \delta(x{-}z) \, ]
\\[4pt]
&&
= 
2 \int_{-L}^L \! dx \, 
j_+^{a_1}(x)\ldots j_+^{a_{m-1}}(x) \, j_0^{d}(x)
\eeaa
where we have again made use of the condition $j\rightarrow 0$ as
$x\rightarrow\pm\infty$ which implies that the 
middle two $\delta$-functions in the penultimate line 
make no contribution.
Thus we find 
\be\label{term2}
\{ \, q_s \, , \, {-1\over2\kappa} 
\int_{-L}^L \! dy \int_{-L}^y dz\,f^{bcd} j_0^c(y)
j_0^d(z) \, \, \} =
- m\, d_{a_1a_2...a_{m-1}c} f^{bcd} 
\int dx \, 
j_+^{a_1}(x)\ldots j_+^{a_{m-1}}(x) \, j_0^{d}(x)
\ee

When we add the terms (\ref{term1}) and (\ref{term2}) we 
can combine $j^d_0 + j^d_1 = j^d_+$.
The symmetrization on tensor indices contracted with $j_+$ currents 
then implies that the total expression 
vanishes, by (\ref{dinv}).

\section{Classical Algebra of Local Charges}

In this section we discuss in detail 
the classical Poisson bracket algebra of
local charges $q_{\pm s}$ of the type (\ref{locch}). 
In calculating these from (\ref{LCPBs}) one finds that the 
terms involving $\delta(x{-}y)$ ({\it i.e.}~the ultra-local terms) 
always vanish by invariance of the $d$-tensors,
leaving just the contributions from the $\delta'(x{-}y)$ terms.
It is clear from (\ref{LCPBs}) that these too are absent if we
are considering charges of opposite chirality, so that 
\be\label{lpb1} 
\{ q_s, q_{-r} \} = 0 \, , \qquad r,s > 0.
\ee 
For charges of the same chirality, however, the result is generally 
non-zero:
\begin{equation}\label{nasty}
\{q_s , q_r \}  =  
(const) \int_{-\infty}^\infty dx\, d_{ca_1 \ldots a_s} 
{d}_{c b_1 \ldots b_{r} } 
j^{a_1}_+ \ldots j^{a_s}_+ 
\del_x ( j^{b_1}_+ \ldots j_+^{b_{r}} ) \,.
\end{equation}
Note that the expression on the right is antisymmetric in $s$ and $r$, 
by integration by parts. 

Before proceeding we must pause to prove our assertion that the 
ultra-local terms do not contribute to (\ref{lpb1}) and (\ref{nasty}).
In (\ref{lpb1}) the ultra-local terms produce 
an integrand which is a combination of 
$$
d_{a_1...a_{s}a}\, d_{b_1...b_{r}b}\; j_+^{a_1} \ldots j_+^{a_{s}}
j_-^{b_1} \ldots j_-^{b_{r}} \, f^{abc} j_\pm^c 
$$
with all fields at the same spacetime argument.
These expressions vanish by invariance (\ref{dinv}) 
of one or other of the $d$-tensors involved.
In (\ref{nasty}) the ultra-local terms contribute 
integrands
$$
d_{a_1...a_{s}a}\, d_{b_1...b_{r}b}\; j_+^{a_1} \ldots j_+^{a_{s}}
j_+^{b_1} \ldots j_+^{b_{r}} \, f^{abc} j_\pm^c \;.
$$
The expression with $j^c_+$ vanishes by invariance of either tensor,
while the expression with $j^c_-$ can be rearranged using invariance 
of the second tensor to yield a result proportional to
$$
d_{a_1...a_{s}a}\, d_{b_1...b_{r}b}\; j_+^{a_1} \ldots j_+^{a_{s}}
j_+^{b_1} \ldots j_-^{b_{r}} \, f^{abc} j_+^c 
$$
and this vanishes by invariance of the first tensor.
This justifies our assertions.

Our aim is to find invariant tensors and conserved currents 
for which the expression (\ref{nasty}) vanishes, so that 
the charges commute.
In the special case $r=1$ and ${d}_{bc} = \delta_{bc}$, 
the integrand in (\ref{nasty}) is clearly a total derivative
and hence the Poisson bracket is zero. This simply means that {\em all\/} the 
local charges (\ref{locch}) commute with energy and momentum:
they are invariant under translations in space and time. 
It is also easy to see that the integrand in (\ref{nasty}) can 
be written as a total derivative if {\em both\/} currents are of the 
form (\ref{confpower}). 
This is a feature of 
any classically conformally-invariant theory whose 
energy-momentum tensor obeys the Virasoro algebra
(\ref{vir}). It is a simple consequence of this that the charges
$\int (T_{++})^n dx$ all commute. 
Finding more interesting sets of commuting charges 
with $s, r > 1$ in the PCM is rather more involved, as we shall see.

\subsection{Algebra of charges for currents $\J_m$}

The natural currents to consider first are $\J_m$ defined by 
(\ref{simple}), in which case (\ref{nasty}) can be written 
\be\label{nicer}
\{q_s , {q}_r \}  = 
(const) \int_{-\infty}^\infty dx\,{\rm Tr}(t^c j_+^s) 
\, \partial_x 
{\rm Tr}(t^c j_+^{r}) \; .
\ee
To simplify this we will use the completeness condition 
$ X = - t^a \, {\rm Tr} (t^a X) $,
which is valid for any $X$ in the Lie algebra $\Lg$. 
This can be applied directly to the integrand in (\ref{nicer}) if we
know that a given power $j_+^r$ or $j_+^s$ lives in 
$\Lg$ for any $j_+$ in $\Lg$. 
Whenever this is true, the completeness condition implies that the 
integrand is proportional to $\del_x {\rm Tr} (j_+^{r+s})$ 
and hence the charges commute.

This argument applies to the orthogonal and symplectic algebras
$so(\ell)$ and $sp(\ell)$, so that in these PCMs the  
charges $\int \J_m dx$ always commute.
To elaborate on this, consider the orthogonal case.
If $X$ is in $so(\ell)$ it is a real anti-symmetric matrix, 
and if $r$ is odd, $X^r$ will also be real and anti-symmetric, 
and hence also in $so(\ell)$.
This implies firstly that $\J_{r+1}$ is zero unless $r$ is odd,
and secondly that under these circumstances we can apply the 
completeness condition 
to write the integrand in (\ref{nicer}) as a total derivative, 
proportional to $\del_x (\J_{r+s})$.
For the symplectic algebras it is easy to see that 
that if $X$ satisfies the defining conditions (\ref{SOUP}) then 
so does $X^r$ if $r$ is odd, so the argument works in precisely the
same way.

Although we are concerned in this paper with PCMs based on simple, 
compact classical groups, it is worth mentioning in passing 
the non-simple case $\Lg = u(\ell)$.
If $X$ is in $u(\ell)$, then so
is $X^r$ if $r$ is odd, or $iX^r$ if $r$ is even.
Either way, the argument applies in just the same way and the charges
$\int \J_{s+1} dx$  commute.\footnote{The same argument also applies
to the case $\Lg = gl(\ell)$ \cite{dickey}, although this is perhaps 
of less direct physical interest since the algebra is non-compact.}
For $\Lg = su(\ell)$ the situation is more complicated, however.
In this case the completeness condition holds only for traceless
matrices $X$ and this property is of course spoiled by taking powers. 
We can still use the completeness condition, but we must do so in a
less direct way. 
Since the generators $t_c$ in (\ref{nicer}) are traceless, we can 
first replace $j_+^r$ by the traceless quantity 
$j_+^r - (1/\ell) {\rm Tr} (j_+^r) $
and then apply the completeness condition to find 
\be\label{n-su}
\{ q_s , q_r \} = (const)
\int_{-\infty}^\infty dx\,{\rm Tr}(j_+^s) \,\partial_x 
{\rm Tr}(j_+^{r}) 
\ee
This is certainly non-zero in general, and so for $su(\ell)$ the charges
$\int \J_m \, dx$ do not commute.
Notice that the non-vanishing bracket is nevertheless 
a conserved quantity that we recognize, namely a differential
polynomial in the currents $\J_m$.

\subsection{Commuting charges for the $su(\ell)$ model}

It is natural to ask whether we can find more
complicated functions of the quantities $\J_m$ which will 
yield commuting charges for the $su(\ell)$ theory. 
To investigate this we need to know the exact form of their Poisson
brackets. It is convenient to set $\kappa =1/2$, which we assume henceforth.
{}From (\ref{PBs}), and making appropriate use of the 
completeness condition as above, we find: 
\begin{eqnarray}
\{ \J_m (x) , \J_n (y) \} = - mn \, \J_{m+n-2}(x) \, \delta'(x{-}y)  
+ {mn \over \ell} \, \J_{m-1}(x) \, \J_{n-1}(x) \, \delta'(x{-}y) 
\nonumber\\   
- {mn(n{-}1) \over n{+}m{-}2} \, \J'_{m+n-2}(x) \, \delta(x{-}y) 
+ {mn \over \ell} \, \J_{m-1} (x) \, \J'_{n-1}(x) \, \delta(x{-}y) 
\label{suPBs}
\end{eqnarray}
We can use these results to search systematically 
for conserved currents $\K_{s+1} (\J_m)$ of homogeneous spin which will 
give commuting charges.
After some laborious calculations we find the following 
expressions for the first few values of the spin: 
\begin{eqnarray}
\K_2 & = & \J_2 
\nonumber\\
\K_3 & = & \J_3 
\nonumber\\
\K_4 & = & \J_4 - {3 \over 2\ell} \, \J_2^2 
\nonumber\\
\K_5 & = & \J_5 - {10\over 3\ell} \, \J_3\J_2 
\nonumber\\
\K_6 & = & \J_6 - {5\over 3\ell} \, \J_3^2 
- {15\over 4\ell} \, \J_4\J_2 + {25\over 8\ell^2} \, \J_2^3 
\label{suN}
\end{eqnarray}
These are the unique combinations (up to overall constants) for which 
the corresponding charges commute.

We can extrapolate from these examples to a general formula.
To construct a charge of spin $s$ we must define a current 
of spin $s{+}1$. In the Lie 
algebra $su(\ell)$ there is (up to an overall constant) a unique 
polynomial in the currents 
$\J_2, \J_3 , \ldots , \J_\ell , \J_{\ell+1}$ 
which is homogeneous of spin $\ell{+}1$ and
which vanishes; we write this 
\be
{\cal A}_{\ell+1} ( \, \J_2 , \J_3 , \ldots , \J_\ell , \J_{\ell+1} \, ) = 0 
\quad {\rm in} \quad su(\ell) \; .
\ee
{}From this we define a current of spin $s{+}1$ for 
$\Lg = su(\ell)$, by the formula 
\be\label{suK}
\K_{s+1} (\, \J_m \, ) = {\cal A}_{s+1} ( \, s \alpha \, \J_m \, )
\qquad {\rm where} \qquad
\alpha = {1 \over h} = {1 \over \ell} \; .
\ee
It is easily checked that (\ref{suK}) reproduces the first 
five examples listed in (\ref{suN}) for any value of $\ell$.

We will prove below 
that the charges $\int \K_{s+1} \, d x$ 
always commute.
Another important fact about the formula (\ref{suK}) is that 
when $s=\ell$ the current vanishes, by construction. 
In fact this also happens whenever $s$ is a multiple of $\ell$,
which we shall also prove below.
Thus, we claim that the formula defines a series of currents whose 
charges can have spin $s$ equal to any integer which is non-zero mod 
$h = \ell$.
To prove these claims, we need to develop some technology. 

Following \cite{balog90} we consider the generating functions 
$A(x , \lambda)$ and $F (x , \lambda)$ defined by 
\be\label{Agen}
A (x , \lambda) = \exp F(x, \lambda) 
= \det ( 1 - \lambda j_+ (x) ) 
\ee
which implies 
\be\label{Fgen}
F(x , \lambda) = {\rm Tr} \log (1 - \lambda j_+ (x) ) 
= - \sum_{r=2}^\infty {\lambda^r \over  r} \J_r (x) \; .
\ee
Observe that $A(x , \lambda)$ is a polynomial in $\lambda$ of degree $\ell$,
with the coefficient of the highest power being 
$(-1)^\ell {\rm det} (j_+)$.
On substituting the series expansion for $F(x,\lambda)$ into (\ref{Agen}),
we obtain non-trivial identities satisfied by the $\J_m$
as the coefficients of $\lambda^r$ must vanish for 
$r > \ell$ (for details, see {\it e.g.}~\cite{AJM98}). 
In particular, the polynomial
${\cal A}_{\ell +1}$ introduced above is the coefficient of
$\lambda^{\ell+1}$. 
Our definition (\ref{suK}) can now be re-written
\be\label{sudef}
{\cal K}_{s+1} = A(x,\lambda)^{s / h} \; {\Big |}_{ \lambda^{s+1} } 
= \exp {s\over h} F(x, \lambda) \; {\Big |}_{ \lambda^{s+1} }  
= \exp \left ( - {s\over h} \sum_{r=2}^\infty 
{\lambda^r \over r } \J_r \right ) \; {\Big |}_{ \lambda^{s+1} } 
\ee

If $s = p h$ with $p$ an integer, then $A(x,\lambda)^{s/h}$ is a 
a polynomial of degree $p h = s$.
The current $\K_{s+1}$ then vanishes.
If $s/h$ is not an integer, however, then $A(x,\lambda)^{s/h}$ will 
be a power series in $\lambda$ with infinitely many terms,
each with a non-vanishing coefficient in general.
This shows that the conserved currents and their charges 
exist precisely when the spin $s$ is non-zero mod $h = \ell$.

Finally we are in a position to prove that the 
charges we have defined commute.
This is done by first calculating the Poisson brackets for the generating
functions and then extracting the desired charges as the coefficients
of particular terms:
\be
\{ q_s , q_r \} = \int \! dx \int \! dy \, \{ A(x,\mu)^{s/h} , \,
A(y,\nu)^{r/h} \} \quad {\Big |}_{\mu^{s+1} \, \nu^{r+1} } \; .
\ee 
In the equations that follow, we will suppress the arguments of 
fields only when there is no possible ambiguity.

{}From (\ref{suPBs}) it can be shown (after some effort) that
\bea 
\{ F(x, \mu) , F(y, \nu) \} \; = \; 
\mu^2 \nu^2 
\left [ \,
{\del_\mu F(x,\mu) {-} \del_\nu F(x,\nu) \over \mu{-}\nu} + 
{ \del_\mu F (x,\mu) \del_\nu F (x,\nu) \over h} \,
\right ] \delta'(x{-}y)
\phantom{XX}
\\
+ \; \mu^2 \nu^2 
\left [ \, {F(x,\mu)'{-} F(x,\nu)' \over (\mu{-}\nu)^2} 
\, - \,{\del_\nu F(x,\nu)' \over \mu{-}\nu} 
\, + \,{\del_\mu F (x,\mu) \del_\nu F (x,\nu)' \over h} \, \right ]
\delta (x{-}y) 
\nonumber
\eea
and from this it follows that\footnote{The corresponding formula 
in \cite{balog90}, eqn.~(4.21), 
seems to contain a misprint; it is not antisymmetric
under the interchange $x \leftrightarrow y$ and $\mu \leftrightarrow \nu$.}
\bea 
\{ A(x, \mu) , A(y, \nu) \} \; = \; 
\mu^2 \nu^2  
\left [ \, {1 \over \mu{-}\nu}(\del_\mu{-}\del_\nu) + 
{\del_\mu \del_\nu \over h} \, \right ] 
A(x,\mu) \, A(x,\nu) \, \delta'(x{-}y)
\phantom{XXXX} \\
+ \; \mu^2 \nu^2 \left [ \, {1\over \mu{-}\nu}(\del_\mu{-}\del_\nu) + 
{\del_\mu \del_\nu \over h} \, \right ] A(x,\mu) \, A(x,\nu)' \, \delta(x{-}y)
\nonumber 
\phantom{XX} \\
+ \; 
{\mu^2 \nu^2 \over (\mu{-}\nu)^2} \left [ 
\, A(x,\mu)' A(x,\nu) - A(x,\mu)A(x,\nu)' \,
\right ]
\delta (x{-}y) \, .
\nonumber
\eea
It can be checked that this is antisymmetric under exchanging 
$x \leftrightarrow y$ and $\mu \leftrightarrow \nu$.
Next we compute 
\bea 
\mbox{\hskip -5pt} 
{1 \over pq \, \mu^2 \nu^2} \{ A(x, \mu)^p , A(y, \nu)^q \} \; = \;
\left [ \, {1 \over \mu{-}\nu} \left \{ {\del_\mu\over p} - {\del_\nu
\over q} \right \} 
+ {1 \over h} { \del_\mu \del_\nu \over pq } \, \right ] 
A(x,\mu)^p A(x,\nu)^q \, \delta'(x{-}y)
\nonumber \\
+ \; \left [ \, {1 \over \mu{-}\nu} \left \{ {\del_\mu \over p}- {\del_\nu
\over q} \right \} 
+ {1 \over h} {\del_\mu \del_\nu \over pq} \, \right ] 
A(x,\mu)^p (A(x,\nu)^q)' \, \delta(x{-}y)
\nonumber \\
+ \; {1 \over (\mu{-}\nu)^2} 
\left [ \, {1\over p} (A(x,\mu)^p)' A(x,\nu)^q - {1\over q} A(x,\mu)^p 
(A(x, \nu)^q)' \, \right ] 
\delta (x{-}y)
\nonumber
\eea
and this implies 
\be
\int \!dx \int \!dy \; \{ A(x, \mu)^p , A(y, \nu)^q \} = 
pq \, \mu^2 \nu^2 \int \! dx \, 
\left [ 
\left\{ {\del_\mu \over p}- {\del_\nu \over q} \right \} {1 \over \mu{-}\nu} 
+ {1 \over h} {\del_\mu \del_\nu \over pq} \right ] 
A(\mu)^p (A(\nu)^q)' 
\nonumber
\ee
To find the brackets of charges of spins $s$ and $r$ we must 
extract the terms of degree $s{+}1$ in $\mu$ and $r{+}1$ in
$\nu$ from this expression.
Provided one is restricting to just these powers,
this means that in the formula above we can replace 
$\mu \del_\mu \rightarrow s$ and $\nu \del_\nu \rightarrow r$.
The result is 
\beaa 
\{ q_s , q_r \} &=& 
\int dx \int dy \; \{ A(x, \mu)^p , A(y, \nu)^q \} \, \,
\; \; {\Big |}_{\mu^{s+1} \nu^{r+1}} \nonumber\\
&=& pq \, \mu \nu 
\int dx \, 
A(\mu)^p (A(\nu)^q)' \left [ 
\left \{ {s \over p} \nu - {r \over q} \mu \right \} {1 \over \mu{-}\nu} 
+ {1 \over h} {rs \over pq} \right ]
\; \; {\Big |}_{\mu^{s+1} \nu^{r+1}} 
\eeaa
The integrand indeed vanishes when 
$p=s/h$ and $q=r/h$, implying that the charges commute as claimed.

\subsection{More commuting charges for $so(\ell)$ and $sp(\ell)$}

A similar approach allows us to construct more general sets of
commuting charges for PCMs based on other classical Lie groups.
For the orthogonal and symplectic algebras the 
Poisson brackets of the currents $\J_m$ are somewhat simpler
(we again set $\kappa =1 /2$):
\be\label{soPBs} 
\{ \J_m (x) , \J_n (y) \} = - mn \, \J_{m+n-2}(x) \, \delta'(x{-}y) 
- {mn(n{-}1) \over m{+}n{-}2} \, \J'_{m+n-2}(x) \, \delta (x{-}y) 
\ee 
Using these we can again search systematically for 
polynomials $\K_{s+1} (\J_m)$ which produce commuting charges,
at least for some low-lying values of the spin.
This reveals a family of currents similar to (\ref{suN}),
except this time there is a single free parameter $\alpha$ which is 
allowed to take an arbitrary value. The first few examples are: 
\begin{eqnarray}
\K_2 & = & \J_2
\nonumber\\[1pt]
\K_4 & = & \J_4 - {1\over 2} (3\alpha) \, \J_2^2 
\nonumber\\
\K_6 & = & \J_6 - {3\over 4}(5\alpha) \, \J_4 \, \J_2
+{1\over 8} (5\alpha)^2 \J_2^3 
\nonumber\\
\K_8 & = & \J_8 -{2\over 3}(7\alpha) \, \J_6 \, \J_2
-{1\over 4}(7\alpha) \, \J_4^2 +
{1\over 4}(7\alpha)^2 \, \J_4 \, \J_2^2
-{1\over 48} (7\alpha)^3 \, \J_2^4 
\label{soN}
\end{eqnarray}

The polynomials appearing above actually coincide with those of the
same degree in (\ref{suN}) if one replaces $\alpha \rightarrow 1/\ell$, and
if one also takes into account the fact that $\J_m$ 
is non-zero for the orthogonal and symplectic algebras only 
if $m$ is even. This immediately suggests the analogous general definition
\be\label{soK} 
\K_{s+1} ( \, \J_m \, ) = {\cal A}_{s+1} ( \, s \alpha \, \J_m \, )
\ee 
Once again, it can be shown that the resulting charges
$\int \K_{s+1} \, d x$ always commute, this time for any value of the
parameter $\alpha$.
Notice that this new one-parameter family of currents 
interpolates the two simplest families we found previously for the
orthogonal and symplectic algebras. 
When $\alpha \rightarrow 0$ we have 
$\K_{2m} \rightarrow \J_{2m}$ and in the limit 
$\alpha \rightarrow \infty$ we have (with a suitable rescaling)
$\K_{2m} \rightarrow (\J_2)^m$. 

To prove that these new currents give commuting charges we again use
generating functions.
Since $\J_m$ is now non-zero only for $m$ even,
it is convenient to introduce two modified generating
functions 
\be 
B(x,\lambda) = \exp G(x,\lambda)
\ee 
where 
\be
B(x, \lambda ) = A (x , \sqrt{\lambda}) = \det ( 1 - \sqrt{\lambda} j_+ (x) ) 
\ee
and 
\be
G(x , \lambda) = F(x , \sqrt{\lambda}) = {\rm Tr} \log 
(1 - \sqrt{\lambda} j_+(x)) 
= - \sum_{r=1}^\infty {\lambda^r \over  r} \, \J_{2r} (x) 
\ee
and to express the Poisson brackets in terms of these.
The general definition of the currents for the orthogonal and 
symplectic algebras can then be written
\be\label{sodef}
\K_{s+1} = B(x,\lambda)^{\alpha s} \; |_{\lambda^{(s+1)/2} }
\ee
and we wish to show that 
\be
\{ q_s , q_r \} = \int dx \int dy \, \{ B(x,\mu)^{\alpha s} ,
B(y,\nu)^{\alpha r} \} \quad |_{\mu^{(s+1)/2} \, \nu^{(r+1)/ 2} }
\ee 
vanishes.

{}From (\ref{soPBs}) we find
\bea
\{ G(x, \mu) , G(y, \nu) \} \; = \; 
{4\mu \nu \over \mu{-}\nu} \left [ {\vphantom{{1\over 2}}}
\, \mu\del_\mu G(x,\mu) - \nu\del_\nu G(x,\nu)\, \right ] \delta'(x{-}y) 
\phantom{XXXXXXXX}
\\
+ \; {4 \mu \nu \over \mu{-}\nu}
\left [ \, {1\over2}{\mu{+}\nu\over \mu{-}\nu} ( \, G(x,\mu)'{-} G(x,\nu)') 
- \nu \del_\nu G(x,\nu)' \, \right ] \delta (x{-}y) \, 
\nonumber
\eea
which implies\footnote{
As with the $su(\ell)$ case, the corresponding formula 
in \cite{balog90}, eqn.~(4.22), 
seems to contain a misprint; it is not antisymmetric
under the interchange $x \leftrightarrow y$ and $\mu \leftrightarrow \nu$.}
\bea 
\{ B(x, \mu) , B(y, \nu) \} \; = \; 
{4 \mu \nu \over \mu{-}\nu} 
\left [ \; \vphantom{{1\over2}} 
( \mu \del_\mu - \nu \del_\nu ) B(x,\mu) B(x,\nu) \, \delta'(x{-}y)
\right . 
\phantom{XXXXXX}
\\
+ \; ( \mu \del_\mu - \nu \del_\nu ) B(x,\mu) B(x,\nu)' \, \delta(x{-}y)
\phantom{XXXX}
\nonumber 
\\
+ \; \left . {1\over 2} {\mu{+}\nu \over \mu{-}\nu} \,
( \, B(x,\mu)' B(x,\nu) - B(x,\mu) B(x,\nu)' \, ) \, \delta (x{-}y)
\; \right ]
\nonumber
\eea
Proceeding as before, we then calculate 
\bea 
{1 \over 4pq \, \mu \nu} \{ B(x, \mu)^p , B(y, \nu)^q \} \; = \; 
{1 \over \mu{-}\nu} \left [ {\mu\del_\mu\over p} -
{\nu \del_\nu\over q} \right ] B(x,\mu)^p B(x,\nu)^q \, \delta'(x{-}y)
\phantom{XXXXX} \nonumber \\
+ \; {1 \over \mu{-}\nu} 
\left [ {\mu \del_\mu \over p}- {\nu \del_\nu \over q} \right ]
B(x,\mu)^p (B(x,\nu)^q)' \, \delta(x{-}y)
\phantom{XXX} \nonumber \\
+ \, {1\over 2} {\mu{+}\nu \over (\mu{-}\nu)^2}
\left [ \, 
{1\over p} (B(x,\mu)^p)' B(x,\nu)^q - {1\over q} B(x,\mu)^p (B(x,\nu)^q)' 
\, \right ] 
\delta (x{-}y)
\nonumber
\eea
and hence
\be
\int \!dx \! \int \!dy \, \{ B(x, \mu)^p , B(y, \nu)^q \} = 
2 pq \mu \nu \! \int \! dx \, 
\left [ \,
{ 2 \mu \del_\mu {+} 1 \over p} \, - \, {2 \nu \del_\nu {+} 1 \over q} 
\, \right ] {1 \over \mu{-}\nu} 
B(\mu)^p (B(\nu)^q)' 
\ee
When we extract the coefficients of $\mu^{(s+1)/ 2}$ and
$\nu^{(r+1)/2}$ we can replace 
$\mu \del_\mu \rightarrow {(s{-}1)/ 2}$ and 
$\nu \del_\nu \rightarrow {(r{-}1)/2}$ in the expression above.
The integrand is then proportional to 
\be
{pq } \, {\mu \nu \over \mu{-}\nu} 
\left [ {s\over p} - {r\over q} \right ] B(\mu)^p (B(\nu)^q)'
\ee
This clearly vanishes if $p= \alpha s$ and $q= \alpha r$ for any 
$\alpha$, as claimed. 

\subsection{The Pfaffian charge and its generalizations}

We have shown that any PCM based on a classical algebra has 
infinitely many commuting local charges constructed from combinations
of invariant tensors of type (\ref{strace}). 
Moreover, these charges come in sequences,
each associated with an exponent of the algebra, or a
{\em primitive\/} invariant tensor, and with the spins in each sequence 
equal to this exponent modulo the Coxeter number, $h$. 
But there is one primitive invariant
tensor which is not of the type (\ref{strace}) and which has therefore
been absent from our discussion so far---this is the Pfaffian
(\ref{pfaff}) with its associated current (\ref{pfaffcurr}). 
It is natural to expect that our results can be extended so as 
to include this last invariant; we now show how this can be done.

We first investigate the behaviour of the 
Pfaffian charge in $so(2\ell)$, 
with current $\p_\ell$, with respect to the other local charges which
we have constructed for this algebra, with currents given in (\ref{sodef}). 
The relevant Poisson brackets are:
\begin{eqnarray} 
&&\mbox{\hskip -18pt}
\{ \p_\ell (x) , \J_{2m}(y) \} = - m \, \p_\ell(x) \, \J_{2m-2} \, 
\delta'(x{-}y) - 
m {2m{-}1\over 2m{-}2} \, \p_\ell (x) \, \J'_{2m-2}(x) \, \delta (x{-}y) 
\qquad (m \neq 2) \nonumber\\[2pt]
&&\mbox{\hskip -18pt} 
\{ \p_\ell (x) , \J_2(y) \} = -2\ell \, \p_\ell (x) \, \delta' (x{-}y)
- 2\p'_\ell (x)
\, \delta(x{-}y) \; . \label{pfPB}
\end{eqnarray}
A derivation of these is described in an appendix.
By calculating the brackets directly, we find that
for the first few examples listed in
(\ref{soN}), the charges $\int \K_m dx$ 
commute with the Pfaffian charge $\int \p_\ell dx$ provided 
we choose $\alpha = 1/h$, where $h = 2\ell{-}2$ is the Coxeter number of 
$so(2\ell)$. By considering the bracket of the Pfaffian current with the 
generating function $B(x,\lambda)$, this can be extended to a 
proof of commutation of the Pfaffian charge 
with all the charges $\int \K_m dx$ defined by (\ref{soN}).
We omit the details, however, in favour of a more complete
treatment from an alternative point of view, as follows.

We are concerned not just with the Pfaffian current but also with
finding generalizations 
$\p_{\ell + ah}$ for all integers $a \geq 0$ (where 
the subscript denotes the spin, as usual).
In other words, we expect it to be 
just the first member of a sequence of conserved quantities 
whose spins repeat modulo the Coxeter number.
It is far from clear {\it a priori\/} how these generalizations 
should be defined, but the answer is, rather remarkably, 
already contained in the generating functions we have been considering.
We have shown how to use the generating functions 
to define the currents $\K_m$ as coefficients in
an expansion in {\em ascending\/} powers of $\lambda$.
It turns out that the Pfaffian and its generalizations naturally
emerge from a similar expansion in {\em descending\/} powers of
$\lambda$.
The formula is then exactly the one already given in (\ref{sodef})
but with $\alpha = 1/h$: 
\be
\p_{\ell + a h} = B(x, \lambda)^{ (a+1/2)} |_{\lambda^{ a(\ell-1)
+\ell/2} }
\qquad a = 0, 1, 2 , \ldots \quad .
\ee
The choice $\alpha = 1/h$ is made so that we recover the Pfaffian 
current when $a=0$.

To illustrate how this works it is best to consider 
the simplest possible example: $\Lg = so(6)$ with $\ell
=3$ and $h = 4$ in the discussion above.
Now 
\be 
B (x ,\lambda) = \exp G(x,\lambda) = {\rm det }( 1 - \sqrt{\lambda} j_+ )
= 1 + \lambda Q_2 + \lambda^2 Q_4   + \lambda^3 \p_3^2
\ee
where for convenience we have defined
\be
Q_2 = -{1 \over 2} \, \J_2 \qquad {\rm and} \qquad  
Q_4 = -{1 \over 4} \, ( \, \J_4 - {1\over 2} \J_2^2\, ) \; .
\ee
To expand in descending powers of $\lambda$ we write
\be
B(x, \lambda)^p = \lambda^ {3p} \, \p_3^{2p} 
\left (\, 1 + {1\over \lambda} Q_4
\p_3^{-2} + {1 \over \lambda^2} Q_2 \p_3^{-2} + {1 \over \lambda^3}
\p_3^{-2} \, \right )^p
\ee
and then expand the bracket to any desired order using the binomial
theorem. 
This gives the following results (up to overall constants)
for the first few generalizations of the Pfaffian charge:
\begin{eqnarray}
\p_7 \, \, & = & \p_3 \, Q_4 
\nonumber\\
\p_{11} & = & \p_3 \, Q_4^2 + {4 \over 3} \, \p_3^3 \, Q_2 
\nonumber\\
\p_{15} & = & \p_3 \, Q_4^3 + 4 \p_3^3 \, Q_4 \, Q_2 + {8\over 5} \, \p_3^5 
\nonumber\\
\p_{19} & = & \p_3 \, Q_4^4 + 8 \p_3^3 \, Q^2 _4 \, Q_2 + {32\over 5} \, \p_3^5
\, Q_4 + {16\over 5} \, \p_3^5 \, Q_2^2 
\label{pfcharges}
\end{eqnarray}
Since $so(6) = su(4)$ we can compare these results with the
non-trivial odd-spin currents predicted by (\ref{sudef}).
Taking due account of normalizations arising from
the inequivalence of the defining representations, 
we find exact agreement (details are given in an appendix). 

Returning now to the general case, we re-iterate the important point
that our definitions of the 
currents $\K_m$ and $\p_m$ are both given by the equation
(\ref{sodef}), but with the understanding that the 
expansions are carried out in ascending and descending powers
respectively. Our proof that (\ref{sodef}) led to commuting charges
did not involve any {\em explicit\/} expansion in the parameter 
$\lambda$; the information about the power of $\lambda$ to be extracted 
was used only to replace the homogeneous differential
operators $\mu \del_\mu$ and $\nu \del_\nu$ by appropriate 
integers. Consequently, our arguments apply equally well if one or
both of the charges considered is of Pfaffian type.
Thus they ensure that the sets 
$\int \K_m dx$ and $\int \p_m dx$ all commute with one another
when $\alpha = 1/h$. 

Finally, it is natural to ask whether new charges could be constructed
for the other classical series $a_\ell$, $b_\ell$ and $c_\ell$ by 
considering expansions in descending powers in a similar way.
If we expand $A(x,\lambda)^{s/h}$ in descending powers, it is not
difficult to see that there is a term of order $\lambda^{s+1}$ only if the 
degree of $A(x,\lambda)$ as a polynomial in $\lambda$ exceeds 
the Coxeter number $h$. 
For the algebras $a_\ell$, $b_\ell$ and $c_\ell$ the degree of 
$A(x,\lambda)$ is less than or equal to the Coxeter number,
so that no non-trivial charges can be defined in this way.
The special feature of $d_\ell$ in this respect 
is that $A(x,\lambda)$ is of degree $2\ell$ which exceeds 
$h = 2 \ell{-}2$, so the construction of charges
through an expansion in descending powers works only in this case.

\section{Comments on the quantum PCM}

It would be very interesting to know whether the classical 
charges we have discussed are still conserved in the quantum theory,
and whether or not the classical Poisson brackets we have shown to vanish
receive corrections when they are elevated to quantum commutators. 
Unfortunately the subtle relationship between the classical PCM
lagrangian and its corresponding quantum theory makes these
questions extremely difficult to tackle.
In this section we give a brief summary of some relevant facts about
the quantum PCM and explain how these fit together with our 
classical results.

The non-local charges are conserved in the quantum theory.
Their behaviour has been studied \cite{berna91,lusch78} 
using a point-splitting regularization of $Q^{(1)a}$ 
and they were found to obey a quantum version of the classical 
Yangian symmetry $Y_L(\Lg) \times Y_R(\Lg)$. An important novel 
feature is that the Poisson bracket of the Lorentz boost generator 
with $Q^{(1)}$, which vanishes classically, develops a term at 
order $\hbar^2$. This is essential to the construction of the 
quantum S-matrices, which would otherwise
be trivial.
Point-splitting regularization would be a very much more 
complicated procedure for the local charges, 
since they involve products of many 
currents taken at a single point. Another approach which is likely
to be rather cumbersome would be to regularize the model on a
lattice. To our knowledge, neither of these approaches has been
developed to study the quantum behaviour of the local charges.
In the absence of a tractable explicit quantization procedure,
the only information comes from less direct arguments.\footnote{
Although the calculations have not been carried
out, it would be surprising if the vanishing 
commutators between the non-local and local charges received 
quantum modifications, since it is difficult to see 
how the known S-matrices could be consistent if the local charges did
not commute with the Yangian.}

\subsection{Goldschmidt-Witten anomaly counting}

In the method of Goldschmidt and Witten \cite{gold80,witt97} 
(see also \cite{aref78}) one considers all possible quantum
anomaly terms which might spoil a classical conservation equation 
$\del_- J = 0$. The form of this equation
reflects the classical conformal invariance of the theory.
One cannot expect it to survive unscathed in the quantum theory, where
conformal invariance is broken, but if the right-hand side
gets a quantum correction which can be written as a derivative,
then we will still have a conserved quantity,
albeit of a modified form. 
The argument is reminiscent of Zamolodchikov's approach
\cite{zam89} to perturbed conformal field theory, though it
pre-dates it. 

To be more precise, suppose we have linearly-independent 
conservation equations of the form $\del_- J_i = 0$ with $i=1, \ldots
, n$ which have a certain prescribed behaviour under all 
global symmetries of the theory. 
The only quantum modifications which can appear on the
right-hand side are operators with the same mass dimension and the 
same behaviour under continuous and discrete symmetries.
Let $A_i$ with $i = 1 , \ldots , p$, be a linearly-independent set of
such operators.
We can also enumerate the linearly-independent 
total-derivative terms $B_i$ with $i = 1 , \ldots , q$, which again 
have the same behaviour under all symmetries.\footnote{
In deciding which $A$s are independent, we are free to use the
classical equations of motion, because any quantum modifications
appearing in the Heisenberg equations will correspond to operators 
with the correct dimensions and invariance properties to ensure that 
they will already occur in our list.
Similarly, in counting the $B$s, as explained in \cite{gold80}.
}
Since each of the $B$s is expressible 
as a combination of $A$s, we have $q\leq p$. Now 
if $n - p + q > 0 $, then there are at least this many 
combinations of the classical conservation equations which survive in
the quantum theory, because this is the number of linearly-independent
combinations for which the right-hand side is guaranteed to be a 
spacetime divergence.

Goldschmidt and Witten wrote down lists of 
$A$s and $B$s for conserved quantities in the PCMs as functions of
the field $g$. We have found it much more convenient to use the 
Lie algebra-valued currents $j_\mu$, particularly in settling
the all-important question of which $A$s and $B$s are independent.
We shall use only the $L$ currents for definiteness, dropping the label $L$
henceforth. 
It is important to keep careful track of the behaviour of these 
currents and their derivatives under the discrete symmetries of the PCM.
{}From (\ref{parity}) we see that under $\g$-parity 
$ \pi : j_+ \mapsto -g^{-1}  j_+ g $ and 
$j_{++}\equiv \partial_+ j_{+} \mapsto 
-g^{-1} \,j_{++}\, g$, but the situation is more
complicated for higher derivatives.
To overcome this we introduce quantities 
\be
j_{+++}\equiv \partial_+ j_{++}\, -\, {1\over 2\kappa} [j_+,j_{++}] \,,
\qquad
j_{++++}\equiv \partial_+ j_{+++}\, -\, {1\over 2\kappa} [j_+,j_{+++}] \,,
\quad \ldots
\ee
which can easily be shown to have the following 
simple behaviour under all discrete symmetries
\begin{eqnarray}
\pi \;: &&
j_{++\ldots+} \mapsto -g^{-1} \,j_{++\ldots+}\, g \nonumber \\[0.1in]
\gamma \;: &&
j_{++\ldots+} \mapsto j^*_{++\ldots+} = -j_{++\ldots+}^T \\[0.1in]
\sigma \;: &&  j_{++\ldots+} \mapsto M j_{++\ldots+} M^{-1}\qquad 
\Lg =so(2\ell) \,.
\end{eqnarray}
Note that for $SO(2\ell)$ all the currents are even under $\sigma$
except for the spin-$\ell$ Pfaffian current, which is odd.

The first example of a conserved current is $\J_2 = {\rm Tr} (j_+^2)$,
the energy-momentum tensor, which we certainly expect to survive
quantization. Indeed, there is only one possible anomaly
$A_1=$Tr$(j_- j_{++})$, only one derivative 
$B_1=\partial_+$Tr$(j_-j_+)$, and in fact $A_1 = B_1$.
This modification of the original conservation law 
reflects the non-vanishing of the trace of the energy-momentum tensor 
quantum mechanically, corresponding to the breaking of conformal symmetry.
The next example is $\J_3 = {\rm Tr}( j_+^3)$, which is non-trivial only
for $SU(\ell)$. This current is odd under both $\pi$ and $\gamma$.
Here again there is one anomaly, $A_1 = {\rm Tr}(j_{++}\{j_-,j_+\})$,
and one derivative $B_1 = \partial_+{\rm Tr}(j_-j_+^2) $, with $A_1 =
B_1$; the conservation again survives quantization.

The case of currents with spin 4 is the most interesting.
There are two classical conserved currents, 
$\J_4 = {\rm Tr} (j^4_+)$, and $\J_2^2 = ({\rm Tr} (j^2_+))^2$.
These are even under each of the discrete symmetries. 
We find
\[
\begin{array}{lll}
A_1 = {\rm Tr}(j_-j_{++++}) & 
\hspace{1in}& B_1=\partial_+{\rm Tr}(j_-j_{+++}) \\[0.1in]
A_2 = {\rm Tr}(j_-j_+){\rm Tr}(j_+j_{++}) 
&& B_2=\partial_+ \left(
{\rm Tr}(j_-j_+){\rm Tr}(j_+^2) \right) \\[0.1in]
A_3 = {\rm Tr}(j_-j_{++}){\rm Tr}(j_+^2) 
&& B_3=\partial_+{\rm Tr}(j_-j_+^3)
\\[0.1in]
A_4 = {\rm Tr}(j_+^2\{j_-,j_{++}\}) 
&& B_4=\partial_-{\rm Tr}(j_{++}^2) \\[0.1in]
A_5 = {\rm Tr}(j_-j_+j_{++}j_+) && 
\end{array}
\]
(Note that another apparent possibility among the $B$s,
$\partial_-{\rm Tr}(j_+j_{+++})$, is not allowed; it is not independent of
$B_3$.) All other possibilities are disallowed: for $SU(\ell)$
these terms are odd under $\gamma$, while for other groups
the traces vanish.
Thus in all cases we have $p=5$, $q=4$ but the number of classical 
currents is $n=2$.
We conclude that there is at least one linear combination
of ${\rm Tr}(j_+^4)$ and $({\rm Tr}(j_+^2))^2$
which will survive quantization.\footnote{
We have reached the same conclusions as \cite{gold80} 
regarding the existence of spin-3 and spin-4 currents.
In comparing our lists of anomalies and derivatives with theirs, 
however, we should point out some discrepancies. There seem to be 
misprints and/or errors in eqns.~(19) and (23) of \cite{gold80}:
the terms $A_4$ in (19) and $A_2$ in (23) do not have the correct
behaviour under discrete symmetries. Furthermore, the terms $B_1$ and
$B_2$ in eqn.~(24) of \cite{gold80} are not independent, since they can be
related using the equations of motion. Any obvious modification of the
term $A_2$ in eqn.~(23) to give it the correct symmetry 
can similarly be related to $A_1$, confirming
our counting above of one anomaly and one derivative for the spin-3
current, rather than two of each, as claimed in \cite{gold80}.
This underscores our opinion that it is clearest to work with quantities
valued in the Lie algebra -- {\it i.e.\/} currents $j_\mu$ rather than
the field $g$.}

There is only one other instance in which the counting arguments are
known to ensure a quantum conservation law. 
It was shown in \cite{gold80} that conservation of 
the Pfaffian current in $SO(2\ell)$
always generalizes to the quantum theory; we shall not reproduce the 
details here.
For all other currents which we have investigated the
Goldschmidt-Witten method is inconclusive.\footnote{ 
For spin-5 currents in the $SU(\ell)$ PCM which are odd under 
$\pi$ and $\gamma$ we find $n=2$, $p=8$ and $q=6$.
For spin-6 currents which are even under $\pi$ and $\gamma$ 
we find $n=5$, $p=25$, $q=18$, for $SU(\ell)$, and 
$n=4$, $p=23$, $q=17$ for $SO(\ell)$ or $Sp(\ell)$.}
We should emphasize however that these counting criteria,
although sometimes {\em sufficient\/} to show quantum conservation, are 
never {\em necessary\/}. The fact that the counting fails should
certainly not be interpreted as meaning that there is no quantum
version of a given classical conservation law, but rather that these
simple arguments are insufficient to settle the issue either way.
Moreover, we have seen that the arguments guarantee the existence of at
least one higher-spin conserved charge in each PCM, which 
is believed to be sufficient for integrability and 
factorization of the $S$-matrix \cite{parke80}.

Note also that the counting arguments, when successful, 
give no information about {\em which combinations\/} of the classical 
currents might survive. 
{}From our earlier work on the classical Poisson brackets of these
charges, we have found for the $a_\ell$ and $d_\ell$ series that there were 
unique preferred sets of charges which all commute with one another. 
It is natural to anticipate that these are 
the combinations which generalize to the quantum theory.
For the $b_\ell$ and $c_\ell$ series, however, such charges are 
not unique and we have no means of discriminating 
amongst the possibilities.

\subsection{S-matrices, particle multiplets and discrete symmetries}

In the exact S-matrix approach to the PCMs, the particle states are
assumed to lie in representations $(V,\bar V)$ of the global symmetry
group $\g_L \times \g_R$ \cite{abda84,ogie86}. The 
representation $V$ of $\Lg$ always contains an irreducible component 
$V_i$ which is one of the fundamental representations (associated to a
node $i$ of the Dynkin diagram) of $\Lg$.
For the $a_\ell$ and $c_\ell$ series the representations are exactly 
$V = V_i$, but for the $b_\ell$ and $d_\ell$ algebras $V$ may be reducible
in general.
It is actually more natural to regard $V$ as acted on by the
entire Yangian $Y(\Lg)$, with $(V,\bar V)$ a representation 
of $Y_L \times Y_R$;
then $V$ is precisely one of the fundamental representations of the 
Yangian $Y(\Lg)$ \cite{ogie86,chari91,kleber96}.
Based on this assignment of representations, the full $S$-matrices have
been determined for the $a_\ell$ and $c_\ell$ PCMs, while 
for the $b_\ell$ and $d_\ell$ models the scattering amongst the vector
and spinor particles has been found (as well as some amplitudes
involving second-rank tensor particles---in principle all other amplitudes 
are determined by the bootstrap procedure, but these have not been 
calculated).

The action of charge conjugation on $Y_L \times Y_R$ 
representations is $\gamma: (V,W) \mapsto (\bar V, \bar W)$,
while the effect of $\g$-parity is to exchange $\g_L$ and $\g_R$, so that 
$ \pi : (V , W)\mapsto (W , V)$. 
Both discrete symmetries map a particle multiplet $(V, \bar V)$ to
itself if $V\ \cong \bar V$.  
But if $V \not\cong \bar V$ then an implication of either symmetry is
that mass-degenerate Yangian representations 
$(V, \bar V)$ and $(\bar V, V)$ must appear together in the
spectrum, as proposed in \cite{ogie86}. 

We have shown that the local charges commute with the Yangian classically.
We will now assume that the same holds in the quantum theory,
so that each local charge takes a constant value on a particle multiplet
$(V,\bar V)$. We want to show how this is compatible with the
assignment of representations and the effects of the discrete symmetries.
Recall in particular that $\g$-parity 
leaves invariant, or commutes with, a local charge whose spin $s$ is an odd
integer, but reverses the value of, or anticommutes with, a local charge for 
which $s$ is an even integer.

The algebras $b_\ell$, $c_\ell$ and $d_{2\ell}$ have only real
representations, $V \cong \bar V$.
A related fact is that the exponents are always odd integers, and so the 
associated local charges commute with $\g$-parity.
This is certainly consistent with the representation 
content $V \otimes V$, which is the simplest kind of multiplet 
whose states can be $\g$-parity singlets.

For the algebras $a_\ell$ and $d_{2\ell+1}$, however,
there is always one exponent which is an even integer, 
and hence at least one local charge $q$ which anti-commutes with 
$\g$-parity.
The only way to have a simultaneous eigenstate of $q$ and
$\pi$ is for the eigenvalue of $q$ to vanish, and so 
on particle multiplets of the form $(V , V)$ we must have $q=0$.
Conversely, Yangian representations on which 
$q\neq 0$ must appear in $\g$-parity pairs.
The fact that the algebra has an even exponent is linked to
the occurrence of complex representations, $V \not\cong \bar V$,
and the multiplets 
$(V , \bar{V})$ and $(\bar{V}, V)$ can indeed be organized into 
$\g$-parity doublets provided they are eigenspaces of 
$q$ with opposite eigenvalues.
Precisely the same phenomenon occurs in 
ATFTs, where it is the even-spin charges which enable states
to be distinguished from their mass-degenerate conjugates \cite{brade90}.

To complete the discussion, we consider the special 
case $d_{\ell} = so(2\ell)$ and the additional discrete symmetry
$\sigma$ which exchanges the spinor representations $S^\pm$.
The representations $S^{\pm}$ are real for $\ell$ even and 
complex for $\ell$ odd.
For $\ell$ even the Pfaffian charge commutes with $\pi$, but
anti-commutes with $\sigma$.
The particle multiplets are $(S^+, S^+)$ and $(S^-, S^-)$ which are 
eigenstates of $\pi$, but these representations are exchanged by
$\sigma$, so its eigenstates lie in $(S^+, S^+) \,\oplus\, (S^-, S^-)$. 
For $\ell$ odd, the Pfaffian charge anti-commutes with both $\pi$ and 
$\sigma$. 
Now the particle multiplets are $(S^+, S^-)$ and $(S^-, S^+)$, and the
eigenstates of both $\pi$ and $\sigma$ lie in $(S^+, S^-)\,\oplus\, 
(S^-, S^+)$.

\subsection{Dorey's Rule}

The occurrence in affine Toda theories (ATFTs) of local 
conserved charges with spins equal to the exponents of the 
underlying Lie algebra modulo the Coxeter number 
leads to the elegant rule for particle fusings discovered
by Dorey \cite{dorey91}.
In ATFTs these particle fusings appear both in the tree-level
three-point couplings (which can be found easily from the classical
lagrangian) and also in the exact S-matrices. For PCMs the
corresponding three-point fusings are defined solely  
by the S-matrices with their Yangian symmetry.
Nevertheless, it has been proved \cite{chari95}, by exploiting 
the technology of Yangian representations in a highly non-trivial 
fashion, that Dorey's rule applies to PCM particle fusings too.

We have constructed a set of local conserved charges in each classical
PCM with exactly the same patterns of spins as those appearing in ATFTs.
We are not at present able to prove that these {\em all\/} 
survive quantization (though some certainly do),
but this fact would provide a very natural explanation of the validity
of Dorey's rule for PCM S-matrices, at least for the simply-laced algebras.
For the non-simply-laced cases there are some additional subtleties,
which we now briefly discuss.

The non-simply-laced ATFTs appear in dual pairs, typically involving
an untwisted and a twisted algebra. Let us consider the example of 
the pair of algebras $c_\ell^{(1)}$ and $d_{\ell+1}^{(2)}$.
The charges have spins equal to the 
exponents of $c_\ell$, but their values depend on the coupling constant:
in the weak-coupling limit they have the 
$c_\ell^{(1)}$ tree-level values, 
while in the strong-coupling limit they are associated in a similar way
with $d_{\ell+1}^{(2)}$.
Dorey's construction does not allow for a coupling constant
dependence in the values of the local charges, 
and gives the tree-level couplings either for $c_\ell$, 
the set of which we shall call $D(c_\ell)$, 
or (when suitably generalized \cite{dorey92}) for $d_{\ell+1}^{(2)}$, 
which we shall call $D(d_{\ell+1}^{(2)})$. It is then the intersection
of the two sets, $D'(c_\ell)\equiv D(c_\ell) \cap D(d_{\ell+1}^{(2)})$, 
which gives the correct fusings for the bootstrap principle applied to
the quantum ATFT S-matrices.

The $c_\ell$ PCM S-matrices \cite{mack92b} also have $D'(c_\ell)$ 
fusings. But in the quantum PCM there is no coupling
constant dependence: the classical coupling is replaced by the overall
quantum mass-scale (dimensional transmutation).
The $c_\ell$ PCM mass ratios are actually those
of the tree-level $d_{\ell+1}^{(2)}$ ATFT, whilst
the values of the other conserved charges must similarly be fixed,
with no coupling-dependence.
The outstanding issue is whether it is the values taken by 
these local charges which are sufficient to restrict
the PCM fusings to $D'(c_\ell)$ (rather than,
say, $D(d_{\ell+1}^{(2)})$), or whether some more subtle 
restriction is taking place.

\section{Summary and Conclusions}

We have carried out a rather thorough investigation of local charges 
in principal chiral models. We have shown that any local charge 
constructed from a symmetric invariant tensor commutes with the
non-local Yangian charges. We have studied the algebra of these
local charges amongst themselves and found that for each
classical algebra there is a commuting family with spins equal to the
exponents modulo the Coxeter number. 
These are defined by the universal formula for the currents 
\be
{\cal K}_{s+1} = A(x,\lambda)^{s \alpha} \; {\Big |}_{ \lambda^{s+1} } 
\qquad {\rm where} \qquad A(x,\lambda) = \det (1 - \lambda j_+) 
\ee
and $\alpha =1/h$ (or more generally for the $b_\ell$ and $c_\ell$
series, $\alpha$ can be arbitrary).
This formula also defines a current associated with the Pfaffian
invariant for the $d_\ell$ algebras, 
as well as generalizations of this current, provided 
we consider an expansion in descending rather than ascending 
powers of $\lambda$.

The existence of infinitely many conserved charges in involution is
a pre-requisite for the classical integrability of any field theory. 
It is worth emphasizing that the brackets (\ref{PBs}) do not apparently
allow such a construction 
via the $r$-matrix formalism of classical inverse
scattering.\footnote{The quantum current algebra is a different matter:
in the quantization method developed by Faddeev and 
Reshetikhin \cite{fadd86} for instance 
there is a manifestly Lorentz-covariant quantum 
current algebra derived from a spin model, with spin-scale $S$.
The classical limit $\hbar\rightarrow 0$ and 
the continuum limit $S\rightarrow\infty$ do not commute. 
If taken in one order, the result is (\ref{PBs}), but if the order of 
the limits is reversed, one finds a different set of classical
Poisson brackets without non-ultralocal terms, which {\it could\/} 
lead to charges in involution via an $r$-matrix.}
We have shown explicitly how the Poisson brackets (\ref{PBs}) 
{\em can\/} lead to infinite sets of charges in involution,
irrespective of the
applicability of the classical $r$-matrix formalism.

Let us now turn to some broader questions surrounding the 
co-existence of local and non-local charges in integrable field
theories, and how these can lead to common conclusions 
via very different chains of argument.
It is worth emphasizing that this happens at the most fundamental
level. The celebrated results which constrain the structure of
the S-matrix in two \cite{parke80} (and four \cite{cole67}) dimensions
make use of the local character of conserved quantities in an essential way.
Such arguments cannot be applied to non-local charges, and yet these still
lead to factorized S-matrices 
through their quasitriangular Hopf algebra (or quantum group) structure.
In the PCMs, the local charges appear to offer a more 
natural explanation of why the fusings follow Dorey's rule than is 
currently available from Yangian representations. 
It would be interesting to investigate other integrable models with 
such results in mind, and to determine whether features usually 
associated with the many exotic properties of non-local charges might also be
explained by the existence of local charges, perhaps more simply.

It is also natural to seek some general,
model-independent way of relating local and non-local conserved quantities. 
The Yangian $Y(\Lg)$ has a trivial centre \cite{chari94}, so 
it seems that we cannot hope to construct local charges 
which commute as `Casimirs' in $Y(\Lg)$ by taking polynomials or 
series in the Yangian charges. 
Another avenue would be to consider the transfer matrix 
which generates the non-local charges, but it is far from clear to us
how our local charges might emerge from this (see also
{\it e.g.}~\cite{chered79,ogiel}). 
{}From yet another point of view, there are hints of a connection of the
type we seek in recent work of Frenkel and
Reshetikhin on deformed $W$-algebras (\cite{frenkel97}
and references therein).

Finally we mention that many of the features we have identified
in this paper also appear in the supersymmetric principal 
chiral model, which we shall deal with in a forthcoming paper
(some preliminary results are reported in \cite{EHMM1}).

{\bf Acknowledgments}

We thank Jose Azc\'arraga, Patrick Dorey and G\'erard Watts for discussions.
JME is grateful to PPARC (UK) for an Advanced Fellowship.
NJM thanks Pembroke College Cambridge for a Stokes Fellowship, during
which much of this work was carried out. 
MH is grateful to St. John's College, Cambridge for a Studentship.
AJM thanks the Royal Commission of 1851 for a Research Fellowship.

\vspace{0.2in}
\section{Appendix: An alternative derivation of Poisson brackets}

The PCM can be regarded as a special case of a 
general $\sigma$-model with lagrangian
$$
{\cal L} = {1 \over 2} \, g_{ij} (\phi) \, \del_\mu \phi^i \, \del^\mu \phi^j
$$
where $\phi^i$ are coordinates on some target manifold with metric
$g_{ij} (\phi)$. The momenta conjugate to the fields $\phi^i$ are
$$
\pi_i = {\del {\cal L} \over \del \dot \phi^i} = g_{ij} (\phi) \, \dot
\phi^j 
$$
with the standard non-vanishing equal-time PBs
$$
\{ \, \phi^i (x) , \, \pi_j (y) \, \} = \delta^i{}_j \, \delta (x-y)
\, .
$$

We may now consider a current of the form
$$
J^a_\mu = E^a_i \, \del_\mu \phi^i
$$
where $E^a_i (\phi)$ are vielbeins on the target manifold satisfying
$$
E^a_i E^a_j = g_{ij}
$$
(Whether these currents are actually conserved or not is irrelevant
for the arguments here.)
In terms of the canonical coordinates $\phi^i$ and $\pi_i$ we have
$$
J^a_0 = E^a_i g^{ij} \pi_j \qquad J^a_1 = E^a_i \phi'{}^{\, i} \, . 
$$
The PB algebra of these currents can now be calculated routinely,
although the general result requires some effort and is not 
particularly illuminating.

Important simplification occurs for the special case of a group
manifold, with currents defined by the vielbeins
$$ 
E^{R\, a}_i = {\rm Tr}(t^a g^{-1} \del_i g) \, ,
\qquad E^{L \, a}_i = - {\rm Tr}(t^a \del_i g g^{-1})
$$ 
where 
$$ 
E^R_i = - g^{-1} \del_i g \, , \qquad 
E^L_i = \del_i g g^{-1} 
$$ 
are the left-invariant and right-invariant forms on the group
respectively (so our labels $L$ and $R$ signify the symmetries under
which the vielbeins {\em transform}). To simplify the current algebra
calculations it is necessary to use only the properties
$$ 
\del_{[i} E_{j]} = E_{[i}E_{j]}
$$ 
(the Maurer-Cartan relations) which are easily verified from the
definitions above.
The Poisson brackets (\ref{PBs}) then follow.

\section{Appendix: Computing Poisson brackets in $SO(2\ell)$ PCM}

The following method applies to any of the 
local conserved charges in the
$SO(2 \ell)$ PCM, including the Pfaffian and its generalizations.  

At each point in space,
$j_+(x)$ is a real $2 \ell {\times} 2 \ell$ 
antisymmetric matrix. Let its skew eigenvalues be
$\lambda_i$, so that there exists an orthogonal $U$ and block diagonal
$D$ with   
$$
U j_+ U^{-1} = D = \sum_{i=1}^{\ell} 
\lambda_i M_i = {\rm diag} \left ( 
\pmatrix{0&\lambda_1\cr-\lambda_1&0\cr}
\ldots \pmatrix{0&\lambda_\ell\cr-\lambda_\ell&0\cr} \right ) \, .
$$
The block diagonal matrices $\{M_i\}$ form a basis for the 
Cartan subalgebra $so(2) \oplus \ldots \oplus so(2)$ 
of $so(2\ell)$, and are clearly normalized 
so that ${\rm Tr}(M_i M_j) = -2 \delta_{ij}$.

Any function of the current $j_+(x)$ which is invariant under the 
adjoint action of the Lie algebra must depend only on the 
eigenvalues $\lambda_i$. For example, we have 
$$
\J_{2n} = {\rm Tr} (j^{2n}_+) = 2 (-1)^n \sum_{i=1}^\ell 
\lambda_i^{2n} \qquad \mbox{and} \qquad 
\p_\ell = \lambda_1 \ldots \lambda_\ell \; .
$$
To compute the Poisson bracket of two invariants 
$P(j_+(x))$ and $Q(j_+(y))$ it therefore suffices to 
know the Poisson brackets between the eigenvalues. 
There are potential complications from the fact that 
the orthogonal matrix $U$ needed to diagonalize the current $j_+(x)$ 
is itself a complicated function of this current.
However, this turns out to be irrelevant to the computation of the
Poisson brackets, as we now show.

From (\ref{LCPBs}) (with $\kappa =1/2$) 
and the arguments of section 4, we know that the only 
contribution to the Poisson bracket is 
$$
\{ P(j_+(x)),Q(j_+(y))\} = 
\frac{\partial P(x)}{\partial j_+^a(x)} \frac{\partial Q(y)}{\partial
j_+^a(y)} \delta'(x-y).
$$
Since $P$ depends on $j_+$ only through its 
eigenvalues we can write
$$
\frac{\partial P(x)}{\partial j_+^a(x)} = \sum_i \frac{\partial P(x)}{\partial
\lambda_i(x)} \frac{\partial \lambda_i(j_+^a(x))}{\partial j_+^a(x)}
$$
(and similarly for $Q$)
and from the definition of $\lambda_i$ given above we find 
$$
\frac{\partial \lambda_i(j_+^a(x))}{\partial j_+^a(x)} = -\frac{1}{2}
{\rm Tr} \left(M_i U t_a U^{-1}\right)+\frac{1}{2} {\rm Tr} \left(M_i
\left[D, \frac{\partial U}{\partial j_+^a} U^{-1} \right]\right).
$$
But the second term involving the commutator vanishes by antisymmetry of 
the matrix $\partial_{j_+^a} U U^{-1}$ (this belongs to the Lie
algebra $so(2\ell)$) in conjunction with the block-diagonal
structure of $D$. 
Using the completeness condition for the generators $\{t^a\}$ of $so(2\ell)$,
we obtain the simple result
$$
\{ P(j_+(x)),Q(j_+(y))\} = {1\over2}\sum_i \frac{\partial P(x)}{\partial
\lambda_i(x)} \frac{\partial Q(y)}{\partial \lambda_i(y)} \delta'(x-y).
$$

This allows an easy derivation of equations such as (\ref{pfPB}). 
In fact, all the Poisson bracket calculations for the $so(2\ell)$ PCM can be 
performed in this way if the formulas involving the generating
functions are re-expressed in terms of eigenvalues.
This provides an important independent check on our calculations using
generating functions.

\section{Appendix: Comparing currents in $SU(4)$ and $SO(6)$ PCMs} 

For $\Lg = su(4)$ the equation (\ref{Agen}) becomes
$$
A(x,\lambda) = 
\exp ( 1 - {1\over2} \lambda^2 \J_2 - {1\over3} \lambda^3 \J_3 
- {1\over4}\lambda^4 \J_4 ) 
= 1 + \lambda^2 \Qt_2 + \lambda^3 \Qt_3 
+ \lambda^4 ( \Qt_4 + {1\over4} \Qt_2^2 ) 
$$
where it is convenient to introduce the quantities
$$
\Qt_2 = - {1\over 2} \J_2 \; , \qquad 
\Qt_3 = -{1\over 3} \J_3 \; , \qquad
\Qt_4 = -{1\over4} ( \J_4 - {1\over 4} \J_2^2) \; .
$$
The definition (\ref{sudef}) is in this case 
$$
\K_{s+1} = A (x, \lambda)^{s/4} \, |_{s+1}
$$
from which we obtain non-trivial, odd-spin currents:
\begin{eqnarray}
\K_3 \, \, & = & \Qt_3
\nonumber \\ 
\K_7 \, \, & = & \Qt_3 \, \Qt_4 
\nonumber\\
\K_{11} & = & \Qt_3 \, \Qt_4^2 - {1 \over 6} \, \Qt_3^3 \, \Qt_2 
\nonumber\\
\K_{15} & = & \Qt_3 \, \Qt_4^3 - {1\over 2} \Qt_3^3 \, \Qt_4 \, \Qt_2 
- {1\over 40} \, \Qt_3^5 
\nonumber\\
\K_{19} & = & \Qt_3 \, \Qt_4^4 - \Qt_3^3 \, \Qt^2_4 \, \Qt_2 
- {1\over10} \, \Qt_3^5 \, \Qt_4 + {1\over 20} \, \Qt_3^5 \, \Qt_2^2 
\nonumber
\end{eqnarray}
With a standard choice of normalization, the relationships between 
invariants in the four-dimensional and six-dimensional representations of 
$su(4) = so(6)$ are 
$$ 
{\rm Tr}_4 X^2 = {1 \over 2} {\rm Tr}_6 X^2 \; ,
\qquad 
{\rm Tr}_4 X^3 = 3i \,{\rm Pfaff}_6 \, (X)  \; ,
\qquad 
{\rm Tr}_4 X^4 = - {1\over 4} {\rm Tr}_6 X^4 + {3 \over 16} 
({\rm Tr}_6 X^2)^2 \; ,
$$
which imply
$$
\Qt_2 = {1 \over 2} Q_2 \; , \qquad
\Qt_3 = - i \p_3 \; , \qquad
\Qt_4 = - {1 \over 4} Q_4 \; .
$$
On substituting into the expressions for the conserved currents
written above we find agreement with (\ref{pfcharges}) up to overall 
constants.

\pagebreak

\end{document}